\documentclass[12pt]{iopart}
\input{amssym}
\usepackage{epsfig}
\global\arraycolsep=2pt

\newcommand{\be}{\begin{equation}}

\newcommand{\ee}{\end{equation}}
\newcommand{\bea}{\begin{eqnarray}}
\newcommand{\eea}{\end{eqnarray}}
\newcommand{\ben}{\begin{equation*}}
\newcommand{\een}{\end{equation*}}
\newcommand{\bean}{\begin{eqnarray*}}
\newcommand{\eean}{\end{eqnarray*}}

\begin{document}
\title{Dephasing in matter-wave interferometry}
\author{Fernando C. Lombardo and Paula I. Villar}
\address{Departamento de F\'\i sica {\it Juan Jos\'e
Giambiagi}, Facultad de Ciencias Exactas y Naturales, UBA;
Ciudad Universitaria, Pabell\' on I, 1428 Buenos Aires, Argentina}

\ead{lombardo@df.uba.ar}

\begin{abstract}
We review different attempts to show the decoherence process in
double-slit-like experiments both for charged particles (electrons) and
neutral particles with permanent dipole moments. Interference is
studied when electrons or atomic systems are coupled to classical or
quantum electromagnetic fields. The interaction between the
particles and time-dependent fields induces a time-varying
Aharonov phase. Averaging over the phase generates a
suppression of fringe visibility in the interference pattern. We
show that, for suitable experimental conditions, the loss of
contrast for dipoles can be almost as large as the
corresponding one for coherent electrons and therefore, be observed. We 
analyze different trajectories in order to show the dependence of the decoherence
factor with the velocity of the particles.
\end{abstract}



\newcommand{\beq}{\begin{equation}}
\newcommand{\eeq}{\end{equation}}
\newcommand{\dalam}{\nabla^2-\partial_t^2}
\newcommand{\mbf}{\mathbf}
\newcommand{\itm}{\mathit}
\newcommand{\beqa}{\begin{eqnarray}}
\newcommand{\eeqa}{\end{eqnarray}}

\section{Introduction: decoherence and the quantum to classical transition}
\label{transition}

Quantum mechanics is one of the most successful theories in Physics.
It can be applied to describe the behavior of solids, the structure
and operation of DNA, and the properties of
superfluids, just to mention a few practical applications.
Despite all its virtues, quantum mechanics is still a
controversial theory. Its description of the physics phenomena is
often confronted with our perceptions of reality and, sometimes,
leads to predictions that can be considered paradoxical. So, at the
root of this ``uneasiness" with quantum theory, lies the superposition
principle ruling Nature's behavior and the classical everyday world
which seems to infringe this superposition principle.

Usually, classicality is identified with the macroscopic world. However,
there are many examples in which we can realize that this association is
not strictly true:  Josephson junctions or nonclassical squeezed
states with macroscopic numbers of photons.
Therefore, we can not systematically treat macroscopic systems as
classical. Might classicality then be an acquired property of the system?
How can we explain our perception of only one outcome out of the
many possibilities that might be? Why don't we observe quantum
interference effects between macroscopic distinguishable states? Why
do we perceive a reality with different alternatives instead of a
coherent superposition of alternatives?

Macroscopic quantum states are never isolated from their
environments \cite{Zurek:2003}. They are not closed quantum systems, and
therefore, they can not behave according to the unitary quantum mechanical 
rules. Consequently, these so often called ``classical" systems
suffer a loss of quantum coherence that is absorbed by the
environment. This {\it decoherence} destroys quantum interferences.
For our everyday world, the time scale at which the quantum
interferences are destroyed is so small that, in the end, the
observer is able to perceive only one outcome.
As far as we see, decoherence is the main process behind the quantum to
classical transition. Formally, it is the dynamic suppression of the
interference terms induced on subsystems due to the interaction with
an environment.

Closed quantum systems are bound to have an unitary evolution in which
the system's purity is preserved and the superposition principle can be applied.
On the other hand, open quantum systems offer a different scenario. As they are in
interaction with an environment (defined as any set of degrees of freedom
coupled to the system which can entangle its states), a ``degradation" of pure
states into mixtures takes place (sometimes, this degradation may be low). These mixtures 
states will often turn to
be diagonal in the set of ``pointer states" \cite{Zurek:2003} which are selected by
the crucial help of the interaction Hamiltonian. They are stable
subjected to its action, i.e., the interaction between the system
and the environment leaves them unperturbed. That's exactly what makes
them a ``preferred" basis.

Let's take for example an interference experiment. The experiment starts
by the preparation of two electron wave packets in a coherent superposition,
assuming each of the charged particles follows a well defined classical path
($C_1$ and $C_2$, respectively), as
\begin{equation}
\Psi(t=0)=(\varphi_1(x)+\varphi_2(x))\otimes \chi_0(y),
\end{equation}
where $\chi_0(\vec y)$ represents the
initial quantum state of the environment (whose set of coordinates
is denoted by $\vec y$). Due to the interaction between the system and the
environment, the total wave function at a later time $t$ is
\begin{equation}
\psi (t) =  \varphi_1(\vec x,t) \otimes
\chi_1 (\vec y,t) + \varphi_2(\vec x,t) \otimes \chi_2 (\vec y,t).
\label{timet}
\end{equation}
It is easy to note that the electrons states  $\varphi_1$ and
$\varphi_2$ became entangled with two different states of the
environment. Therefore, the probability of finding a particle at a
given position at time $t$ (for example when interference pattern is
examined) is,
\begin{equation} \mbox{Prob}(\vec x,t) = \vert \varphi_1
\vert^2 + \vert \varphi_2\vert^2 + 2 {\rm Re}\left(
\varphi_1\varphi^*_2 \int d^3y ~\chi_1^*(\vec
y,t)\chi_2(\vec y,t)\right).
\label{proba}\end{equation}

Last term in the above expression represents quantum
interferences. We define the overlap factor as $F = \int d^3y
~\chi_1^*(\vec y,t)\chi_2(\vec y,t)$. This factor is responsible
for two separate effects. Its phase generates a shift of the
interference fringes, and its absolute value is responsible for
the decay in the interference fringe contrast. Of course, in
absence of environment the overlap factor is not present in the
interference term. When the two environmental states do not
overlap at all, the final state of the bath identifies the path
the electron followed. There is no uncertainty with respect to the
path. Decoherence appears as soon as the two interfering partial
waves shift the environment into states orthogonal to each other.

The loss of quantum coherence can alternatively be explained by
the effect of the environment over the partial waves, rather than
how the waves affect the environment. As has been noted, when a
static potential $V(x)$ is exerted on one of the partial waves,
this wave acquires a phase, \beq \phi = - \int V[x(t)] dt, \eeq
and therefore, the interference term appears multiplied by a
factor $e^{i\phi}$. This is a possible agent of decoherence. The
effect can be directly related to the statistical character of
$\phi$, in particular in situations where the potential is not
static. Yet more importantly,
 any source of stochastic noise would create a
decaying coefficient. For a general case, $\phi$ is not totally
defined, i.e. it is described by means of a distribution function
$P(\phi )$. From this statistical point of view, the phase can be
written as \beq \langle e^{i\phi}\rangle = \int e^{i\phi} P(\phi
) d\phi.\label{IF} \eeq

In this way, the uncertainty in the phase produces a decaying term
that tends to eliminate the interference pattern. This dephasing is
due to the presence of a noisy environment coupled to the system and
can be also represented by the Feynman-Vernon influence functional
formalism. It is easy to prove that Eq.(\ref{IF}) is the influence
functional generated after integrating out the environmental degrees
of freedom of an open quantum system. Therefore, 
the formal equivalence between the two ways
of studying dephasing was shown in Ref.
\cite{Stern:1990}
\beq \langle e^{i\phi}\rangle = F =  \int d^3y
~ \chi_1^*(\vec y,t)\chi_2(\vec y,t). \label{overlap}\eeq 
The overlap factor $F$ encodes the information about the statistical
nature of noise. Therefore, noise (classical or quantum) makes $F$
less than one, and the goal is to quantify how it slightly destroys
the particle interference pattern.

\section{Double-slit-like interference experiments}

In many cases, the interaction with the environment cannot be
switched off. Quantum electromagnetic field fluctuations are well
known for being responsible for the Casimir forces. Less known
is the role of
these fluctuations in the possible destruction of
electron coherence. Thus, for charged particles or neutral atoms
with dipole moment, the interaction with the electromagnetic field
is crucial because it induces a reduction of fringe
visibility.

In Ref.\cite{Villa:2003} authors have analyzed the influence of a conducting
boundary in the decay of the visibility of interference fringes in a
double slit experiment performed with charged particles (or neutral
particles with dipole moment). They considered the zero point
fluctuations as an external environment. As the presence of
a conducting boundary modifies the properties of the vacuum,
it is not surprising that they could also affect the interference
pattern. They assumed an initial state
$|\Psi(0)\rangle=(|\phi_1\rangle+
|\phi_2\rangle)\otimes|E_0\rangle$, where $|E_0\rangle$ is the
initial (vacuum) state of the field and $|\phi_{1,2}\rangle$ are two
states of the electron localized around the initial point.
Therefore, the probability of finding a particle at a given position
was also given by Eq.(\ref{proba}), but with the overlap factor
defined by $F=\langle E_2(t)| E_1(t)\rangle$, simply the overlap between two
states of the field that arise from the vacuum under the influence
of two different sources (the two electron currents $J_{1,2}^{\mu}$).
Not only did they compute it in the presence of
a conducting boundary, but also did it in vacuum in order to compare
both results. In the end, they showed that the presence of the
conducting plane may produce more decoherence in some cases and less
decoherence in others (in relation to the fringe visibility in the
vacuum case). This difference has to do with the orientation of the
conducting plate and the trajectories of the interfering particles.
Thus, the effect of the boundaries does not have a well defined sign
and may produce either more decoherence or  complete recoherence
(i.e., smaller or higher fringe visibility than in the absence of
the conducting boundary). Nonetheless, even in the case where the
fringe visibility is destroyed the most, the quantification of the
decoherence suffered by the open system scales as $v^2$, where $v$
is the velocity of the particles in the non relativistic limit.
Therefore, its magnitude is too small to be measured in the
laboratory.

In Ref.\cite{Vourdas1:2001}, authors have calculated the effect of the
classical and quantum noise in external nonclassical microwaves on
the phase factor that describes electron interference. 
In the presence of such external nonclassical
electromagnetic fields, the phase factor is a quantum mechanical
operator. Using this phase factor operator, they studied the effect
on mesoscopic Josephson junctures and on time-dependent
Aharonov-Bohm and Aharonov-Casher devices. 
In addition to the presence of quantum noise, one can see the importance of
classical fields in the destruction of the interference pattern as
well. Classical fields' effects seem to be of the same magnitude or
even bigger than the ones generated by nonclassical fields.

A very innovative study of the loss of contrast in the interference
pattern of two electron beams was brought to light by Hsiang $\&$
Ford in \cite{Ford:2004}. They studied the effect of time-varying
electromagnetic fields on electron coherence, including the
statistical origin of the Aharonov-Bohm (AB) \cite{Aharonov:1959}
phase $\phi$. However, they didn't consider $\phi$ neither as coming from quantum
fluctuations nor a time dependent field. They included a random
variable $t_0$, which was defined as the electron emission time.
This variable produces a fluctuating phase, and an average
over it is needed in order to obtain the result of the double-slit
interference experiment. In this simple version of decoherence, the
role of a quantum environment is replaced by a time-dependent
external field which gives a time-varying AB phase. In the end, they
studied the ``dephasing" process and computed the reduction of the
interference oscillations, evaluating the overlap factor given in
Eq.(\ref{overlap}). They considered
the case of a linearly polarized monochromatic electromagnetic
wave, propagating in a direction orthogonal to the plane containing
two electron beams. Surprisingly this time, the effect on the
fringes seems to be sufficiently large to be observable.

In \cite{Lombardo:2005} we followed last idea. We evaluated the overlap
factor for coherent neutral particles with permanent (electric and
magnetic) dipoles that are affected by time-varying external fields (linearly
polarized electromagnetic plane-waves and fields inside a waveguide). This
is the generalization of results of Ref. \cite{Ford:2004} to the case
of Aharonov-Casher (AC) \cite{Casher:1984} in atomic systems, where
coherent dipoles follow a
closed path around an external field. As we will discuss, not all
these effects foresee an observable displacement on the interference
pattern related to the phase shift of the wave function of the
system. Dependence on the velocity of the interfering particle is
crucial. We will show different configurations where the decoherence
factor depends on the particle's speed in a different way.

\section{Aharonov phases in external time dependent classical fields}

The AB phase, known to arise when two coherent
electrons traverse two different paths $C_1$ and $C_2$
in the presence of an electromagnetic field, is ($c=\hbar=1$)
\beq \phi=-e \oint_{
\delta \Omega}  dx_{\nu} A^{\nu}(x), \label{faseAB} \eeq where
$\delta \Omega =C_1-C_2$ is a closed spacetime path. If the electromagnetic
field fluctuations happen in a time scale shorter than the total
time of the experiment, this shift of phase results in a loss of
contrast in the interference fringes. Then, the overlap factor (or
decoherence factor) is given by Eq.(\ref{overlap})
\cite{Stern:1990,Ford:2004}. In that expression, angular brackets denote
either an ensemble of quantum noise or a time average over a
random variable.

The phase shift that two neutral particles with
electric and magnetic dipole moments experience due to a classical
time dependent electromagnetic field is known as the Aharonov-Casher
phase and is defined by \beq \phi=- \oint_{\delta
\Omega} a_{\nu}(x) dx^{\nu}, \label{faseAC} \eeq where
$a_{\nu}(x)=(-\mbf{m} \cdot \mbf{B} -\mbf{d} \cdot \mbf{E}, \mbf{d}
\times \mbf{B} - \mbf{m} \times \mbf{E})$ plays the role of $A_\nu$
in the AB case \cite{Anandan:2000}.

In order to evaluate the integral in Eq.(\ref{faseAC}), we consider
the case of a linearly polarized monochromatic wave of frequency
$\omega$ propagating in the $\hat{y}$ direction, with an electric
and magnetic field in the $\hat{z}$ and $\hat{x}$ direction
respectively (different external fields have been analyzed in
\cite{Lombardo:2005}). We will also assume that the particles' path is
confined to the $\itm{{\hat x}-{\hat z}}$ plane. We can write the
plane wave as $\mbf{E}(x)=E_0 \sin(wt-ky)~\hat{z}$, $\mbf{B}(x)=E_0
\sin(wt-ky)~\hat{x}$ and compute $a_\nu$.

Following Ref.\cite{Ford:2004}, we will assume that the phase $\phi$
depends on a random variable $\xi = \omega t_0$ given by the
emission time of the particles. It is the time $t_0$ at which the
center of a localized wave packet is emitted. When the measuring
time takes longer than the flight time, we will observe a result
which is the temporal average over $t_0$. Thus, $t_0$ is a random
variable by which $\phi$ has to be averaged. We can write the AC
phase $\phi$ as \beq \phi (t_0) =-\oint_{\delta\Omega} {\tilde
a}_\nu \sin (\omega t - k y + \omega t_0) ~ dx^\nu = A \cos (\omega
t_0) + B \sin (\omega t_0), \label{tphase}\eeq where ${\tilde
a}_\nu$ are the spatial components of $a_{\nu}(x)$.

The average over the random phase (generating a classical noise) produces a
decoherence factor
\begin{equation*}
F = \langle e^{i\phi}\rangle = \lim_{T\rightarrow \infty}\frac{1}{2T}
\int_{-T}^{T}dt_0 \exp\left\{i\left[A \cos (\omega t_0) + B \sin (\omega t_0)
\right]\right\}=
J_0(\vert C\vert ),
\end{equation*}
where $J_0$ is the Bessel function. The modulus of
complex number $C = A + i B$ measures
degree of decoherence. The overlap factor $F$ decreases from
one to zero as $\vert C\vert$ varies between zero and the
first zero of $J_0$. For larger values of $\vert C\vert$, the
overlap factor oscillates with decreasing amplitude.

A characteristic feature of the usual  AB and AC effects is that the
phase shift is independent of the velocity of the
particle and there is no force on the particle 
\cite{Berry:1991,APV:1988}. Moreover, the phase shift depends only 
on the topology of the closed spacetime path $\delta \Omega$. Of 
course, these properties are not longer
valid when  the external field is time dependent because the
particle does suffer a net force applied on it. Thus, in order to
analyze the dependence upon the trajectory, we will evaluate
Eq.(\ref{faseAC}) for different paths and will find that the phase's
dependence on the velocity is strongly related to the trajectory the
particles follow.

\subsection{Overlap factors for AC phases}

In this subsection, we will estimate the AC phase acquired by two
neutral particles with electric and magnetic dipole moment when
they follow two different trajectories.
To begin with, we will consider an elliptic path (Fig.1 (a)). Each particle
traverses half ellipse (${\cal
C}_1$ and ${\cal C}_2$) sharing the initial and final point. Therefore, 
the elliptic closed path is symmetric with respect to the particles direction 
of propagation. This is what we will call {\it symmetric} trajectory. The quantity
$|C^d_{\rm ellip}|$ is given by \beq |C^d_{\rm ellip}|= 2 \pi \alpha
E_0 d_y J_1 [\omega \tau] \approx \frac{\sqrt 2\pi\alpha
E_0d_y}{(\omega \tau )^{1/2}} = \sqrt \pi\alpha e E_0 L
\left(\frac{v \lambda}{s'}\right)^{1/2} \label{Cd_ellip}, \eeq
where $\tau$ is the time of flight of the dipoles, $\alpha$ is the
maximun distance between dipoles, $d_y$ the electric dipole moment
in the $\hat{y}$ direction
 and $J_1$ the Bessel function of first order. In
the last term, we have used the asymptotic expansion of the Bessel
function for $\omega \tau \gg 1$ (which is expected for
non-relativistic particles). $L$ is the characteristic length of an
atom with electric dipole $d= e L$ ($L \approx 10^{-9} m$),
$\lambda$ is the wavelength of the plane wave, and
$s'$ is the length traveled by the neutral particles at a speed
$v$ and in a time $\tau$. It is important to note that for this
symmetric trajectory $|C^d_{\rm ellip}|$ scales as $\sqrt v$.

\begin{figure}
\centerline{\psfig{figure=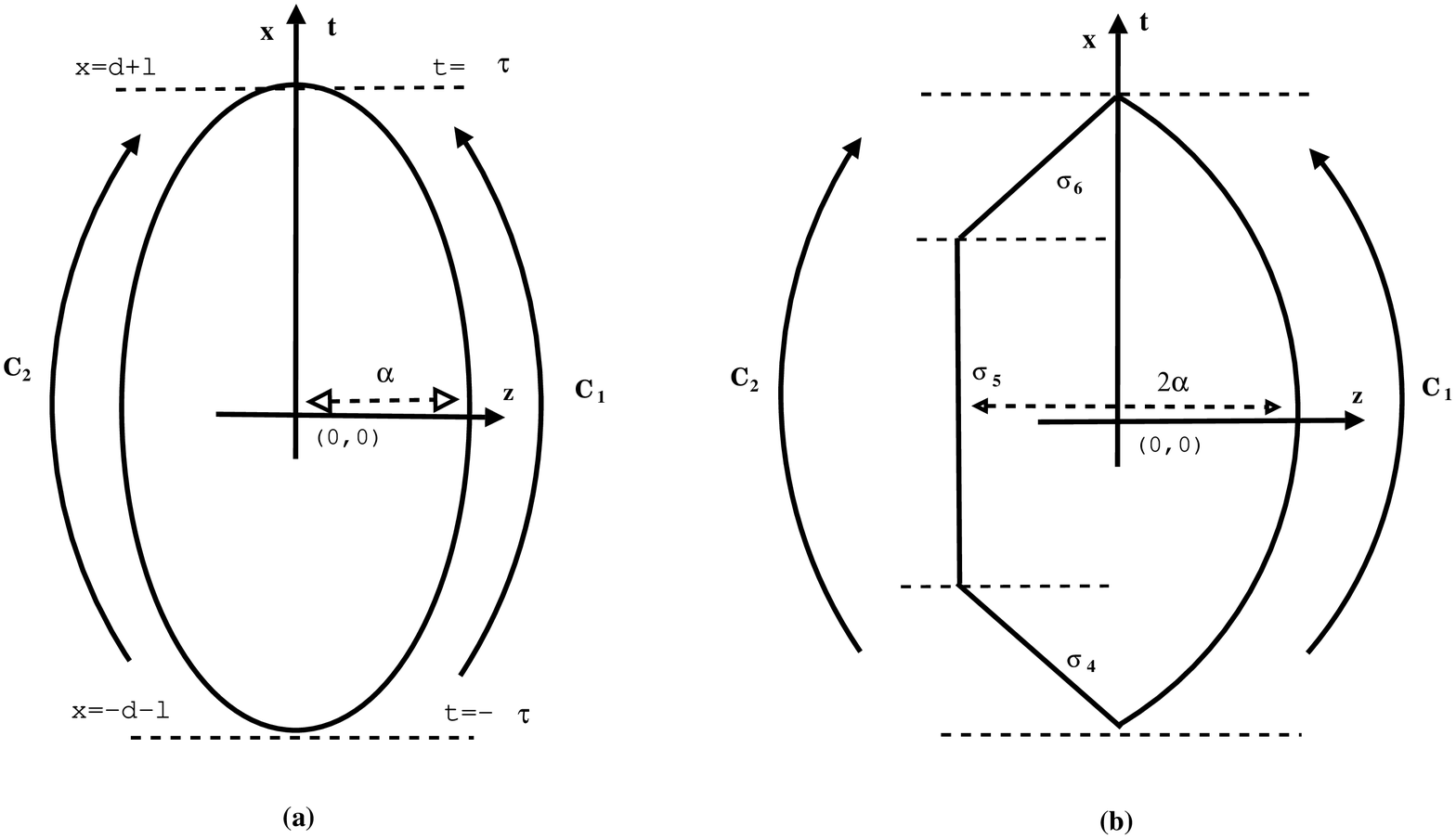,height=6cm,width=9cm,angle=0}}
\caption{Paths ${\cal C}_1$ and ${\cal C}_2$ are shown for the (a) 
elliptic trajectory and (b) the asymmetric one.}
\label{fig1}
\end{figure}

As a second case, we study the decoherence suffered by two neutral
particles that transverse different paths, giving rise to the
asymmetric closed path shown in Fig. 1 (b) (see \cite{Lombardo:2005} for
details)). Velocity dependence will be different from the case of
the symmetric trajectory.

After performing the corresponding integrations in
Eqs.(\ref{faseAC}) and (\ref{tphase}),
$|C^d_{\rm asym}|$ can be approximated (in the slow velocity
limit) by \beq |C^d_{\rm asym}| \approx \frac{e}{\pi}  E_0  L
\lambda\,\, , \label{Cd_asym} \eeq being independent of the
velocity.

\subsection{Overlap factors for AB phases}

It is interesting to check whether the same velocity dependence
applies to the case of the AB phase for charged particles or not.
We will consider the case of electron wave packets traveling
across the above symmetric and asymmetric trajectories. 
Therefore, in order to compare the results for charged
and neutral particles with dipole moments, we will compute the AB
phase for both cases using Eqs.(\ref{faseAB}) and (\ref{tphase}).

If the charged particles traverse the elliptic trajectory, the
quantity $|C^e_{\rm ellip}|$ is \beq |C^e_{\rm ellip}|=  2 \pi \alpha
e E_0 \lambda J_1[\omega \tau] \approx \sqrt {\pi} \alpha e E_0
\lambda \bigg( \frac{v \lambda} {s'} \bigg)^{1/2}, \eeq showing
that the dependence on the velocity is similar for both neutral
and charged particles traveling the same trajectory (although the
velocities differ in magnitude).

The last case we must consider is the one of charged particles
performing the asymmetric trajectory. The decoherence factor can be
now computed as
\beq |C^e_{\rm asym}| \approx \frac{e}{ \sqrt{2\pi}}  E_0 \alpha
\bigg(\frac{v\lambda^3}{s'}\bigg)^{1/2},\label{Ce_asym} \eeq which
depends on the velocity similarly to the preceding case. What is
worthy of note is the fact that while this result does depend on
the velocity of the electrons, the decoherence factor for the
dipoles applied to the same trajectory does not.

\section{Numerical estimations and final remarks}

In \cite{Lombardo:2005}, we have shown that, contrarily to what
might be naively expected, the loss of contrast of the interference
fringes for dipoles might be as big as for the case of electrons.
By the way, if one introduces real numbers to our analytical
results for electrons and dipoles, the results are quite
interesting. In electrons interferometry, the wave packets can be
moved apart up to $100 \mu\rm{m}$ \cite{Hasselbach:1988}. A
typical non-relativistic velocity is $v_e \sim 0.1$. This yields a
relation $\omega \tau \sim 10$ for a field that has a wavelength
of about $100 \mu\rm{m}$. On the other hand, in atomic
interferometry, two neutral particles can be separated up to
$1~\rm{mm}$ \cite{Keith:1991,Pfau:1993}. Typical speeds are of the
order $v_d \sim 10^{-5}$ \cite{Green:1981}.
We will assume an energy flux of 10 $\rm{Watts}/\rm{cm}^2$,
approximately.

With all these values, we can estimate the
$C$ factor for all the cases presented in the previous sections.
The results for electrons are of order one or even bigger, which
means that the effect is experimentally observable. Dipoles'
results are smaller but not that much as one would naively expect.
In the case of the elliptic trajectory, the decoherence factor for
dipoles is $C^d_{\rm ellip}\sim 10^{-3}$, while in the case of the
asymmetric path, one gets $C^d_{\rm asym}\sim 10^{-1}$.

In electron's interference experiments, we have shown that,
in principle, it is possible to obtain a complete destruction of
the interference pattern (setting the value of $\vert C\vert$
equal to a zero of the Bessel function $J_0$). On the other side,
in an interference experiment with dipoles, the best experimental
setup would be the asymmetric trajectory. In this case, the effect
is non negligible thanks to the fact that the $C$ factor is
independent of the velocity. Moreover, if one takes into account
the result in \cite{Lombardo:2005}, one is allowed to increase the
intensity of the external field, since the scattering cross
section for dipoles is much lower than for electrons, being still
possible to neglect the direct interaction with the
electromagnetic field.

Matter-wave interferometry has been largely studied in the last
years. Many theoretical studies has been done around the {\it
mesoscopic} systems \cite{Facchi:2005,Brezger:2002}. ``Mesoscopic" objects are
neither microscopic nor macroscopic. They are generally systems that
can be described by a wave function, yet they are made up of a
significant number of elementary constituents, such as atoms. Famous
examples these days are fullerene molecules $C_{60}$ and $C_{70}$.
Most notably, the quantum interference of these molecules has been
observed \cite{Zeilinger:2003}. In this context, we may estimate the
decoherence factor for dipoles using the experimental values of the
fullerene's experiments. Even though we know it is a toy model for 
the fullerene molecules, it provides a quantitative estimation about the possibility 
of measuring the effect of decoherence for neutral systems. These molecules 
have a speed similar to that of the dipoles we have considered. However, they travel longer
distances. Therefore, it is compulsive to consider a bigger
wavelength, at least of the same order of magnitude that the total
distance the dipoles travel.
It is important to note that, in
the experiment described in \cite{Zeilinger:2003}, a laser beam with a
power of $26\rm{Watts}/\rm{cm}^2$ behaves as an external field.
With all these new values in mind, the decoherence factor for
dipoles when traversing the asymmetric trajectory is $|C_{\rm
asym}| \approx 1$, giving complete destruction of the interference
pattern.

\ack
We would like to thank F.D. Mazzitelli for his collaboration
on this project. We also thanks R. Onofrio for usefull comments, 
and E. Elizalde for the organization of QFEXT'05. This work is supported by
UBA, Conicet, and Fundaci\'on Antorchas, Argentina.

\end{document}